\documentclass{aa}
\usepackage{graphicx}
\usepackage{txfonts}

\begin{document}

\title{A search for microquasar candidates at low galactic latitudes}

\author{J.~M. Paredes\inst{1}
\and M. Rib\'o\inst{1}
\and J. Mart\'{\i}\inst{2}
}

\offprints{J.~M. Paredes, \\ \email{josep@am.ub.es}}

\institute{Departament d'Astronomia i Meteorologia, Universitat de Barcelona, Av. Diagonal 647, 08028 Barcelona, Spain
\and Departamento de F\'{\i}sica, Escuela Polit\'ecnica Superior, Universidad de Ja\'en, Virgen de la Cabeza 2, 23071 Ja\'en, Spain
}

\date{Received 21 May 2002 / Accepted 24 July 2002}

\abstract{
Recent studies of relativistic jet sources in the Galaxy, also known as
microquasars, have been very useful in trying to understand the
accretion/ejection processes that take place near compact objects. However, the
number of sources involved in such studies is still small. In an attempt to
increase the number of known microquasars we have carried out a search for new
Radio Emitting X-ray Binaries (REXBs). These sources are the ones to be
observed later with VLBI techniques to unveil their possible microquasar
nature. To this end, we have performed a cross-identification between the X-ray
ROSAT all sky survey Bright Source Catalog (RBSC) and the radio NRAO VLA Sky
Survey (NVSS) catalogs under very restrictive selection criteria for sources
with $|b|<5\degr$. We have also conducted a deep observational radio and
optical study for six of the selected candidates. At the end of this process
two of the candidates appear to be promising, and deserve additional
observations aimed to confirm their proposed microquasar nature.
}

\maketitle

\section{Introduction} \label{intro}

Microquasars are X-ray binary systems with the ability to generate
relativistic jets. The presence of such collimated features is considered to
be the unmistakable fingerprint of their microquasar nature. The jets are seen
thanks to their synchrotron emission at radio wavelengths, that can be
resolved with modern interferometers. Hence, all microquasars belong to the
class of Radio Emitting X-ray Binaries (REXBs), which contains about $\sim50$
sources among the $\sim280$ known X-ray binaries (Liu et~al. \cite{liu00}; Liu
et~al. \cite{liu01}).

The discovery of microquasars, so far only in our Galaxy, illustrates how
sometimes nature mimics itself at scales orders of magnitude different. The
interesting analogy between quasars and microquasars, both from the
morphological and physical point of view, has been pointed out by several
authors. In addition, microquasars are considered to be good laboratories for
understanding the physics of accretion/ejection processes and strong
gravitational fields (see, e.g., Mirabel \& Rodr\'{\i}guez \cite{mirabel99}). 
The key parameter that distinguishes the quasar/microquasar properties is
believed to be the mass of the compact object in the central engine, varying
over nine orders of magnitude from supermassive to stellar-mass black holes.
The most recent findings of microquasars have been summarized in Castro-Tirado
et~al. (\cite{castro01}).

The current census of microquasar sources in the Galaxy amounts to 14 systems
(Mirabel \& Rodr\'{\i}guez \cite{mirabel99}, Rodr\'{\i}guez \& Mirabel
\cite{rodriguez01}), with \object{LS~5039} (Paredes et~al. \cite{paredes00}),
\object{Cyg~X-1} (Stirling et~al. \cite{stirling01}) and
\object{XTE~J1550$-$564} (Hannikainen et~al. \cite{hannikainen01}) as the most
recent additions. This number could increase dramatically if, as it has been
proposed by Fender \& Hendry (\cite{fender00}), radio emission from X-ray
binaries always arises from relativistic jets. This would imply that all REXBs
are microquasars. However, the situation is not so clear yet and REXBs appear
as a rather heterogeneous group including both high-mass (e.g.,
\object{Cyg~X-3}, \object{LS~I~+61~303}) and low-mass companions (e.g.,
\object{Aql~X-1}, \object{GX~339$-$4}). Moreover, in the latter group, radio
emission has been detected in both atoll and Z-type sources. In addition, the
detection of radio emission does not seem to be limited to a particular type
of compact object, since there are REXBs harboring black holes (e.g.,
\object{GRO~J1655$-$40}, \object{Cyg~X-1}) and neutron stars (e.g.,
\object{Aql~X-1}, \object{Sco~X-1}). Moreover, when one begins to further
divide the group of REXBs according to their optical or compact companion or
whether radio jets are present or not, the number of systems is no longer
enough to perform a meaningful statistical study.

For example, the systems whose compact companion is believed to be a black
hole appear to have jet velocities above $0.9c$, while those with neutron
stars appear to exhibit slower jet flows with speeds below $0.5c$ (Mirabel \&
Rodr\'{\i}guez \cite{mirabel99}). However, Fomalont et~al. (\cite{fomalont01})
have measured speeds up to $0.57c$ in the neutron star system
\object{Sco~X-1}, and a lower limit of {\it only} $0.6c$ has been derived for
the black hole system \object{Cyg~X-1} by Stirling et~al. (\cite{stirling01}),
while systems like \object{Cyg~X-3}, where the nature of the compact object is
unknown, exhibit flows with intermediate speeds of $0.5c$ (Mart\'{\i} et~al.
\cite{marti01}). Therefore, some questions arise: is there a true correlation?
what is the speed limit of the jets in neutron star systems, if any? 

Another interesting issue is the possibility of microquasars being sources of
high energy $\gamma$-rays. Up to now, the most promising candidate is
\object{LS~5039}, which appears to be associated with the third EGRET catalog
(Hartman et~al. \cite{hartman99}) source \object{3EG~J1824$-$1514} (Paredes
et~al. \cite{paredes00}). On the other hand, the 26.5~d periodic REXB
\object{LS~I~+61~303} could be associated with the source
\object{3EG~J0241+6103} (Kniffen et~al. \cite{kniffen97}). Although these
associations will only be confirmed, or discarded, by future missions such as
GLAST, clearly a sample of two sources is a very poor one, and any attempt to
discover new REXBs/microquasars, and look for possible high energy
$\gamma$-ray emission, is warranted.

Hence, the reduced population known gave us a strong motivation to start a
long-term project focused on searching for new microquasars in the Galaxy. In
this context, X-ray, optical and radio catalogs provide a fundamental tool for
the search of new sources with known multiwavelength behavior. Therefore, we
have used the best available X-ray and radio catalogs to search for new REXB
candidates on the basis of positional coincidence, together with very
restrictive selection criteria. As a result we have obtained a sample of 13
radio emitting X-ray sources, which is expected to contain at least some
galactic X-ray binaries. For 6 of the selected candidates, a deep study
involving optical and VLA high resolution observations was carried out in
order to better determine their positions and nature. After these
observations, all but one of the sources have appeared unresolved, even at the
highest VLA resolution, and the positions are known to better than 0.01
arcsecond accuracy. Optical spectroscopy is also currently in progress to
confirm the stellar nature of our candidates, while already carried out VLBI
observations will be reported in a future paper.

In this paper we present the results of the cross-identification together with
radio and optical astrometry and photometry of these sources. We explain the
developed cross-identification method in Sect.~\ref{cross}. We present the
follow-up radio observations in Sect.~\ref{radio}, while the optical ones are
described in Sect.~\ref{optical}. An analysis of the obtained data and the
corresponding discussion are presented in Sect.~\ref{discussion}. We finally
state our conclusions in Sect.~\ref{conclusions}.

\section{Cross-identification between RBSC and NVSS catalogs} \label{cross}

In order to find new microquasar candidates we have to search for new REXBs.
The first step, hence, has consisted of a positional cross-identification of
X-ray and radio sources using the most complete available catalogs at both
wavelengths, namely the ROSAT all sky Bright Source Catalog (RBSC) and the
NRAO VLA Sky Survey (NVSS), which are briefly introduced now.

In the X-ray domain, the RBSC (Voges et~al. \cite{voges99}) contains in its
current version a total of 18\,806 sources in the energy band 0.1--2.4~keV,
and is derived from the ROSAT All Sky Survey (Voges et~al. \cite{voges96}).
Four energy bands in the following ranges of keV are provided: $A$=0.1--0.4,
$B$=0.5--2.0, $C$=0.5--0.9, $D$=0.9--2.0. From these bands, two hardness
ratios are computed: $HR1=(B-A)/(B+A)$, $HR2=(D-C)/(D+C)$. The 1$\sigma$
positional uncertainties are typically in the range 10--20$\arcsec$.

In the radio domain, the NVSS (Condon et~al. \cite{condon98}) catalog covers
the sky north of $\delta=-40\degr$ (82\% of the celestial sphere) at a
frequency of 1.4~GHz (20~cm wavelength) using the VLA configurations D and
DnC. It contains over $1.8\times10^{6}$ sources stronger than its 2.5~mJy
completeness limit. The rms positional uncertainties are less than $1\arcsec$
for sources stronger than 15~mJy and $7\arcsec$ for the faintest detectable
sources. 

Since our aim is to maximize the probability of retaining only galactic REXB
systems by cross-identifying the RBSC and the NVSS catalogs, we have adopted
the following selection criteria:

\begin{enumerate}

\item Sources with absolute galactic latitude $<5\degr$ have been selected
from the RBSC catalog. Since the NVSS has a limit of $\delta>-40\degr$, only
the RBSC sources above this declination have been allowed to continue in the
selection process. Therefore, approximately 75\% of the whole $|b|<5\degr$
area is covered ($l\simeq347$--$259\degr$ and $\alpha\simeq17.2$--$8.6^{\rm
h}$).

\item Among the remaining RBSC sources, those containing screening flags about
nearby sources contaminating measurements or problems with position
determinations (Voges et~al. \cite{voges99}) have been rejected.

\item From statistical RBSC identification studies, Motch et~al.
(\cite{motch98}) concluded that X-ray binaries are essentially recognizable
from their hard X-ray spectra. Therefore, in an attempt to avoid AGNs and
cataclysmic variables, we have only selected the sources with
$HR1+\sigma(HR1)$ higher than 0.9. Although this criterion would exclude
microquasars in an outburst, we note that this situation only lasts a short
time of their life, and the probability of discarding one of them is very low.
A total of 241 RBSC sources remain in our sample at this stage.

\item The X-ray and radio positions must agree within errors. For this
purpose, we have selected NVSS sources within the 2$\sigma$ (95\% probability)
error boxes of the RBSC sources. We have also used the constraint of a maximum
offset of 40$\arcsec$ between the RBSC and the NVSS positions, since, for
higher distances, no reliable identification is expected (Voges et~al.
\cite{voges99}).

\item No extended radio source has been allowed to continue in the selection
process, since any REXB is expected to appear compact at the NVSS
resolution\footnote{The sources resolved only in one axis and with an angular
size smaller than the other axis were allowed to continue on the process.}.

\end{enumerate}
The sources selected with the RBSC/NVSS cross-identification, a total of 35,
were then filtered with complementary optical information using the following
criteria:

\begin{enumerate}

\item We have inspected the SIMBAD database and the NASA/IPAC Extragalactic
Database (NED) for each source, and if it is listed as an extragalactic
object, we have obviously rejected it from the sample.

\item We have also inspected the Digitized Sky Survey, DSS1 and DSS2-red
images\footnote{{\tt http://archive.eso.org/dss/dss/}}, and looked for
possible optical counterparts with position agreement with the NVSS sources.
If an optical object is present and displays extended emission, i.e., with a
galaxy-like appearance, it has also been removed from the sample.

\end{enumerate}
At the end of this process the resulting sample contained 16 sources. Among
them, we found the well known REXB \object{LS~I~+61~303}, which displays a
one-sided jet at milliarcsecond scales (Massi et~al. \cite{massi01}). We also
recovered the well known microquasars \object{LS~5039}, \object{SS~433} and
\object{Cyg~X-3}. It is interesting to note that all these sources are HMXB,
and none of the known microquasars with low mass companions were found by this
technique. This can be easily explained by the fact that most of LMXB are
transients, and the remaining persistent ones are too faint to be present in
the NVSS. Moreover, all but one HMXB persistent microquasars in our explored
range of galactic latitudes, namely \object{LS~5039}, \object{SS~433} and
\object{Cyg~X-3}, have been recovered by our cross-identification method. The
remaining one, namely \object{Cyg~X-1}, although marginally present in the
corresponding NVSS image, does not appear in the NVSS catalog. This was
probably due to the source low flux density and diffuse background around it,
which prevented its automatic detection by the source search software.

After removing the previously known sources of our sample, we ended with a
total of 12 new unidentified REXB candidates. Among them, 7 had offsets
between the X-ray and radio positions within the 1$\sigma$ RBSC position
error, and belong to the hereafter Group~1 sample. The other 5 sources had
offsets between 1--2$\sigma$ and form the Group~2 sample.

When we started this project in 1998 another source fullfiled all the
selection criteria. Now we have performed the cross-identification process
again, in order to present here an updated version of it, and we have found
that this source, namely \object{1RXS~J072418.3$-$071508}, was identified by
Perlman et~al. (\cite{perlman98}) as a quasar (\object{WGA~J0724.3$-$0715} in
their paper and listed as \object{PMN~J0724$-$0715} in the NED database).
Although it is clear that it is not any more a REXB candidate, we have
preferred to include it in the list and present the observational results
obtained up to now. Since the corresponding offset in position for this object
is less than 1$\sigma$, it is included in Group~1.

All the sources belonging to Group~1 (a total of 8), Group~2 (a total of 5)
and the already known REXBs are listed in Table~\ref{crosst}, where Cols.~1 to
4 contain the RBSC names (which also provide the positions), the 1$\sigma$
errors in position, the count rates, and $HR1$. In Col.~5 we show the offsets
between the RBSC and the NVSS positions. In Cols.~6 to 8 we list the NVSS
names (which provide a limited position information), the 1$\sigma$ errors in
position and the NVSS flux densities (at 20~cm wavelength). Finally, galactic
coordinates for all sources are listed in Cols.~9 and 10. The horizontal line
divides the two groups of sources and the already known REXBs.

\begin{figure*}[htpb]
\resizebox{\hsize}{!}{\includegraphics{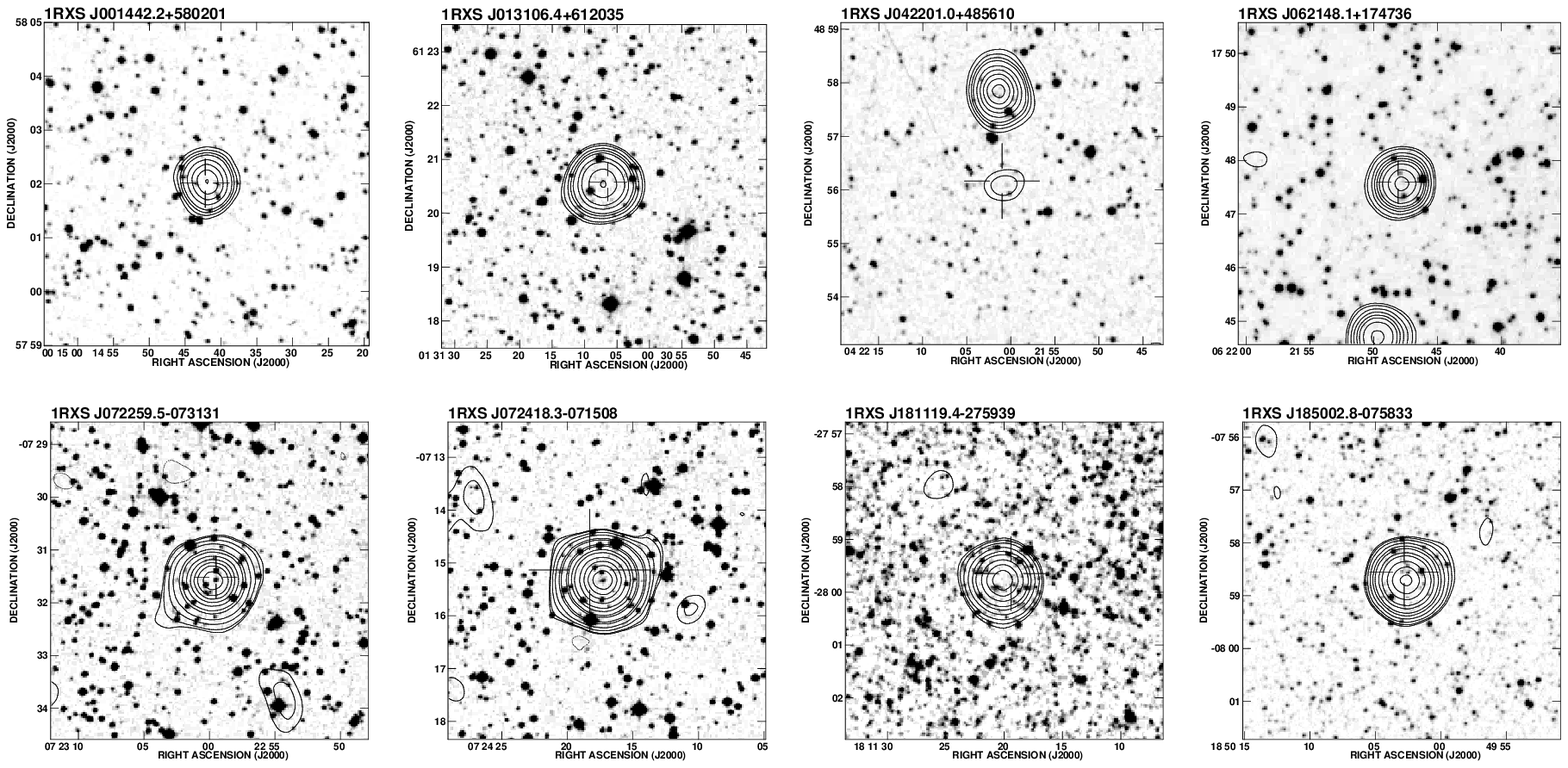}}
\caption{Optical, radio and X-ray composition of Group~1 sources. The NVSS radio contours are overlaid on the $6\arcmin\times6\arcmin$ optical DSS1 images, whereas the open crosses denote the $3\sigma$ RBSC astrometric errors.}
\label{sample}
\end{figure*}

In this paper we will focus on the Group~1 sources, while Group~2 sources will
be eventually studied in the future. In order to have a look at the targets,
we show in Fig.~\ref{sample}, for each source of Group~1, the NVSS radio
contours overlaid on $6\arcmin\times6\arcmin$ optical DSS1 images. To mark the
RBSC positions we have plotted as open crosses the $3\sigma$ errors in
position, allowing a better display than if we had plotted the $1\sigma$
errors.

\begin{table*}
\begin{center}
\caption[]{Selected sources from the RBSC/NVSS cross-identification. Cols.~1 to 4 contain the RBSC name (which also provides the position), the 1$\sigma$ error in position, the count rate, and the hardness ratio 1 for each source. In Col.~5 we show the offset between the RBSC and the NVSS positions, while in Cols.~6 to 8 we list the NVSS name (constructed with truncated coordinates), the 1$\sigma$ error in position and the NVSS flux density for each source. Galactic coordinates for all sources are listed in Cols.~9 and 10. The horizontal lines divide Group~1 (top), Group~2 (middle), and the already known REXB sources (bottom). The source \object{1RXS~J072418.3$-$071508} is a quasar (see text).}
\label{crosst}
\begin{tabular}{l@{~~}c@{~~}c@{~~~~~}ccl@{~~}c@{~~~}c@{~~~~~~~}c@{~~~~}c}
\hline \hline \noalign{\smallskip}
\multicolumn{4}{c}{RBSC} & RBSC/NVSS & \multicolumn{3}{c}{NVSS} & \multicolumn{2}{c}{Gal. coord.}\\
\noalign{\smallskip} \hline \noalign{\smallskip}
1RXS name                  & Pos. err. & Count rate & $HR1$    & Offset      & NVSS name & Pos. err.   & Flux density & $l$ & $b$ \\
               & [$\arcsec$] & [$10^{-2}$~s$^{-1}$] &          & [$\arcsec$] &           & [$\arcsec$] & [mJy]        & [$\degr$] & [$\degr$] \\
\noalign{\smallskip} \hline \noalign{\smallskip}
J001442.2+580201   & ~\,9 & $~\,8.5\pm1.4$ & $1.00\pm0.12$ & ~\,3 & J001441+580202   & 3 & $~~~7.1\pm0.5$  &   118.07 & $-$4.49 \\
J013106.4+612035   & ~\,7 & $  25.0\pm2.4$ & $0.90\pm0.05$ & ~\,6 & J013107+612033   & 1 & $~\,19.1\pm0.7$ &   127.67 & $-$1.16 \\
J042201.0+485610   &   14 & $~\,5.1\pm1.1$ & $1.00\pm0.24$ & ~\,4 & J042200+485607   & 7 & $~~~2.3\pm0.4$  &   154.41 & $-$0.63 \\
J062148.1+174736   & ~\,8 & $~\,8.8\pm1.6$ & $1.00\pm0.13$ & ~\,4 & J062147+174734   & 1 & $~\,12.2\pm0.5$ &   193.78 &   +1.72 \\
J072259.5$-$073131 & ~\,8 & $  17.4\pm2.1$ & $0.94\pm0.05$ & ~\,6 & J072259$-$073135 & 1 & $~\,84.1\pm3.1$ &   223.24 &   +3.52 \\
J072418.3$-$071508 &   23 & $~\,5.2\pm1.2$ & $1.00\pm0.13$ &   19 & J072417$-$071519 & 1 & $330.4\pm9.9$   &   223.15 &   +3.93 \\
~~~~~~~~~~~~~~~~~(Quasar) & & & & & & & & \\
J181119.4$-$275939 &   14 & $  14.0\pm3.1$ & $0.87\pm0.14$ &   13 & J181120$-$275946 & 1 & $~\,27.9\pm1.0$ &  ~~~3.64 & $-$4.43 \\
J185002.8$-$075833 &   13 & $~\,6.1\pm1.6$ & $1.00\pm0.24$ &   10 & J185002$-$075842 & 1 & $~\,49.9\pm1.6$ & ~\,25.66 & $-$3.32 \\
\noalign{\smallskip} \hline \noalign{\smallskip}
J050339.8+451715   &   11 & $~\,6.1\pm1.3$ & $0.85\pm0.17$ &   16 & J050339+451658   & 1 & $~\,34.3\pm1.1$ &   161.81 &   +2.31 \\
J080451.8$-$274924 &   11 & $~\,7.0\pm1.8$ & $1.00\pm0.15$ &   14 & J080451$-$274911 & 1 & $821~~\,\pm25~$ &   245.80 &   +2.01 \\
J082404.3$-$302033 & ~\,8 & $  17.5\pm2.4$ & $0.96\pm0.06$ &   11 & J082403$-$302038 & 1 & $~\,85.0\pm2.6$ &   250.23 &   +4.11 \\
J190333.1+104355   &   14 & $  10.8\pm1.7$ & $1.00\pm0.04$ &   16 & J190333+104409   & 7 & $~~~4.7\pm0.6$  & ~\,43.86 &   +2.22 \\
J205644.3+494011   &   16 & $~\,7.7\pm1.0$ & $1.00\pm0.06$ &   17 & J205642+494005   & 1 & $167.3\pm5.0$   & ~\,89.32 &   +2.76 \\
\noalign{\smallskip} \hline \noalign{\smallskip}
J024033.5+611358   &   13 & $~\,5.4\pm1.1$ & $1.00\pm0.17$ &   19 & J024031+611345   & 1 & $~\,42.2\pm1.3$ &   135.68 &   +1.09 \\
~~~~~~~(\object{LS~I~+61~303}) & & & & & & & & \\
J182615.1$-$145034 &   11 & $~\,6.5\pm1.6$ & $1.00\pm0.16$ &   21 & J182614$-$145054 & 1 & $~\,23.4\pm0.9$ & ~\,16.88 & $-$1.29 \\
~~~~~~~~~~~~~~\,(\object{LS~5039}) & & & & & & & & \\
J191149.7+045857   & ~\,8 & $  52.6\pm3.7$ & $0.96\pm0.03$ & ~\,3 & J191149+045858   & 1 & $867~~\,\pm26~$ & ~\,39.69 & $-$2.24 \\
~~~~~~~~~~~~~~~~\,(\object{SS~433}) & & & & & & & & \\
J203226.2+405725 & ~\,7 & $284.9\pm6.5$~\, & $0.98\pm0.00$ & ~\,7 & J203225+405728   & 1 & $~\,87.3\pm3.2$ & ~\,79.84 &   +0.69 \\
~~~~~~~~~~~~~~(\object{Cyg~X-3}) & & & & & & & & \\
\noalign{\smallskip} \hline
\end{tabular}
\end{center}
\end{table*}

As can be seen in Table~\ref{crosst}, the NVSS errors are much smaller than
the RBSC ones. Hence, the radio positions are more accurate than the X-ray
ones to look for optical counterparts. An inspection of the DSS1 images in
Fig.~\ref{sample} reveals that some candidates have an optical counterpart
approximately in the middle of the radio contours, while others do not. This
is understandable because of the low galactic latitude of these objects, which
implies a high degree of extinction along the line of sight. On the other
hand, crowded fields near the galactic plane also prevent us to be sure about
the counterparts detected. Therefore, the next step of the process is to
obtain accurate radio positions and then search for possible optical
counterparts.

\begin{table*}
\begin{center}
\caption[]{Radio positions, flux densities and spectral indices obtained after the VLA A configuration observations for all sources belonging to Group~1 except \object{1RXS~J181119.4$-$275939} and \object{1RXS~J185002.8$-$075833}. The coordinates are those obtained from the uniform weighted maps of all concatenated 3.6~cm wavelength observations. The flux densities at 3.6 and 6~cm have been obtained from natural weighted self-calibrated maps for each source, with the exception of \object{1RXS~J042201.0+485610}, where self-calibration was not possible.}
\label{radiot}
\begin{tabular}{lrrcccc}
\hline \hline \noalign{\smallskip}
1RXS name             & $\alpha$~(2000) & $\delta$~(2000) & Day & \multicolumn{2}{c}{Flux density [mJy]} & {Spectral index} \\
                      & [h, m, s] & [$\degr$, $\arcmin$, $\arcsec$] & July 1999 & $S_{3.6\,{\rm cm}}$ & $S_{6\,{\rm cm}}$ & $\alpha_{3.6-6\,{\rm cm}}$\\
\noalign{\smallskip} \hline \noalign{\smallskip}
J001442.2+580201         & 00 14 42.1262 &   +58 02 01.219 & ~\,9 &   $6.94\pm0.04$        &   $7.90\pm0.07$        &  $-0.24~\,\pm0.02$~\,\\
                         & $\pm\,0.0013$ &    $\pm\,0.010$ & 22   &   $6.36\pm0.07$        &   $7.05\pm0.09$        &  $-0.19~\,\pm0.03$~\,\\
                         &               &                 & 30   &   $5.98\pm0.06$        &   $6.47\pm0.07$        &  $-0.14~\,\pm0.03$~\,\\
\noalign{\smallskip}
J013106.4+612035         & 01 31 07.2267 &   +61 20 33.376 & ~\,9 &  $16.95\pm0.04$~\,     &  $17.95\pm0.08$~\,     &  $-0.10~\,\pm0.01$~\,\\
                         & $\pm\,0.0014$ &    $\pm\,0.010$ & 22   &  $16.62\pm0.07$~\,     &  $16.49\pm0.09$~\,     &  $+0.01~\,\pm0.01$~\,\\
                         &               &                 & 30   &  $17.43\pm0.07$~\,     &  $17.95\pm0.08$~\,     &  $-0.05~\,\pm0.01$~\,\\
\noalign{\smallskip}
J042201.0+485610         & 04 22 00.5244 &   +48 56 03.634 & ~\,9 &   $0.70\pm0.03$        &   $0.30\pm0.04$        &  $+1.6~~~\pm0.3$~~~~\\
                         & $\pm\,0.0010$ &    $\pm\,0.010$ & 22   &   $0.74\pm0.05$        &   $0.57\pm0.05$        &  $+0.5~~~\pm0.2$~~~~\\
                         &               &                 & 30   &   $0.71\pm0.03$        &   $0.35\pm0.04$        &  $+1.3~~~\pm0.2$~~~~\\
\noalign{\smallskip}
J062148.1+174736         & 06 21 47.7522 &   +17 47 35.078 & ~\,9 &   $9.63\pm0.04$        &   $8.37\pm0.07$        &  $+0.26~\,\pm0.02$~\,\\
                         & $\pm\,0.0007$ &    $\pm\,0.010$ & 22   &   $9.53\pm0.07$        &   $9.87\pm0.07$        &  $-0.06~\,\pm0.02$~\,\\
                         &               &                 & 30   &  $11.45\pm0.05$~\,     &  $10.40\pm0.07$~\,     &  $+0.18~\,\pm0.01$~\,\\
\noalign{\smallskip}
J072259.5$-$073131$\,^a$ & 07 22 59.6809 & $-$07 31 34.805 & ~\,9 &  $40.90\pm0.05$~\,     &  $47.67\pm0.08$~\,     &  $-0.280\pm0.004$\\
                         & $\pm\,0.0007$ &    $\pm\,0.010$ & 22   &  $44.29\pm0.07$~\,     &  $44.5~\,\pm0.1$~~~    &  $-0.009\pm0.005$\\
                         &               &                 & 30   &  $45.37\pm0.08$~\,     &  $54.29\pm0.08$~\,     &  $-0.329\pm0.004$\\
\noalign{\smallskip}
J072418.3$-$071508$\,^b$ & 07 24 17.2912 & $-$07 15 20.339 & ~\,9 & $438.5~\,\pm0.2$~~~~\, & $423.4~\,\pm0.3$~~~~\, &  $+0.064\pm0.002$\\
                         & $\pm\,0.0007$ &    $\pm\,0.010$ & 22   & $455.75\pm0.09$~~~     & $439.3~\,\pm0.2$~~~~\, &  $+0.067\pm0.001$\\
                         &               &                 & 30   & $428.5~\,\pm0.3$~~~~\, & $404.9~\,\pm0.3$~~~~\, &  $+0.104\pm0.002$\\
\noalign{\smallskip} \hline
\end{tabular}
\end{center}
$^a$ Also observed at 20~cm on 1999 July 30, with a flux density of $55.1\pm0.4$~mJy, and showing a one-sided radio jet.\\
$^b$ Also observed at 20~cm on 1999 July 30, with a flux density of $248.3\pm0.9$~mJy.\\
\end{table*}

\section{Radio observations} \label{radio}

The main goal of the radio observations was to obtain accurate sub-arcsecond
positions, but we also wished to monitor the variability of radio flux and
spectrum of our targets. On the other hand, it was also important to
investigate the source structure beyond the NVSS resolution. For REXBs, most
of the source flux density is expected to be concentrated in a compact core
plus possible arcsecond or sub-arcsecond extended features. By observing each
target with the VLA in A configuration, we were able to verify to what extend
our sources were indeed compact, and to look for possible elongations or jets.

To this end, multi-frequency and multi-epoch observations of the Group~1
sources were carried out with the NRAO\footnote{The National Radio Astronomy
Observatory is a facility of the National Science Foundation operated under
cooperative agreement by Associated Universities, Inc.} VLA in A
configuration. We did not observe \object{1RXS~J181119.4$-$275939} and
\object{1RXS~J185002.8$-$075833}, the last two sources of Group~1 in
Table~\ref{crosst}, because they had very different right ascensions, compared
to the other sources, and were not visible during the scheduled VLA observing
time.

Three VLA A configuration observing sessions were carried out on 1999 July 9,
22 and 30 at 3.6~cm (8.4~GHz) and 6~cm (5~GHz) wavelengths (frequencies). A
typical observation consisted of 14 minutes at 3.6~cm and 4 minutes at 6~cm on
each target, preceded and followed by a 2 minute observation of a phase
calibrator. During the last observing session, July 30, we included two
additional 20~cm (1.4~GHz) measurements of \object{1RXS~J072259.5$-$073131}
and \object{1RXS~J072418.3$-$071508}. The amplitude calibrator used in all
cases was \object{3C~48}, while we used the phase calibrator
\object{J0102+584} for the first two sources, \object{J0359+509} and
\object{J0603+177} for the third and fourth sources, respectively, and finally
\object{J0730$-$116} for the last two sources. The data were edited and
calibrated using the AIPS software package of NRAO.

\subsection{Results}

All sources were detected at all frequencies, allowing us to obtain accurate
positions and flux density measurements. To achieve the first goal, we made
use of the phase-reference technique to calibrate the data at 3.6~cm
wavelength, the ones with the highest angular resolution. Then we concatenated
all these data for each source, with the task DBCON within AIPS. Finally, we
performed uniform weighted maps from which positions were obtained after
fitting elliptical Gaussians using JMFIT within AIPS. A realistic estimate of
the position error is about $0.01\arcsec$. 

\begin{figure*}[htpb]
\resizebox{\hsize}{!}{\includegraphics{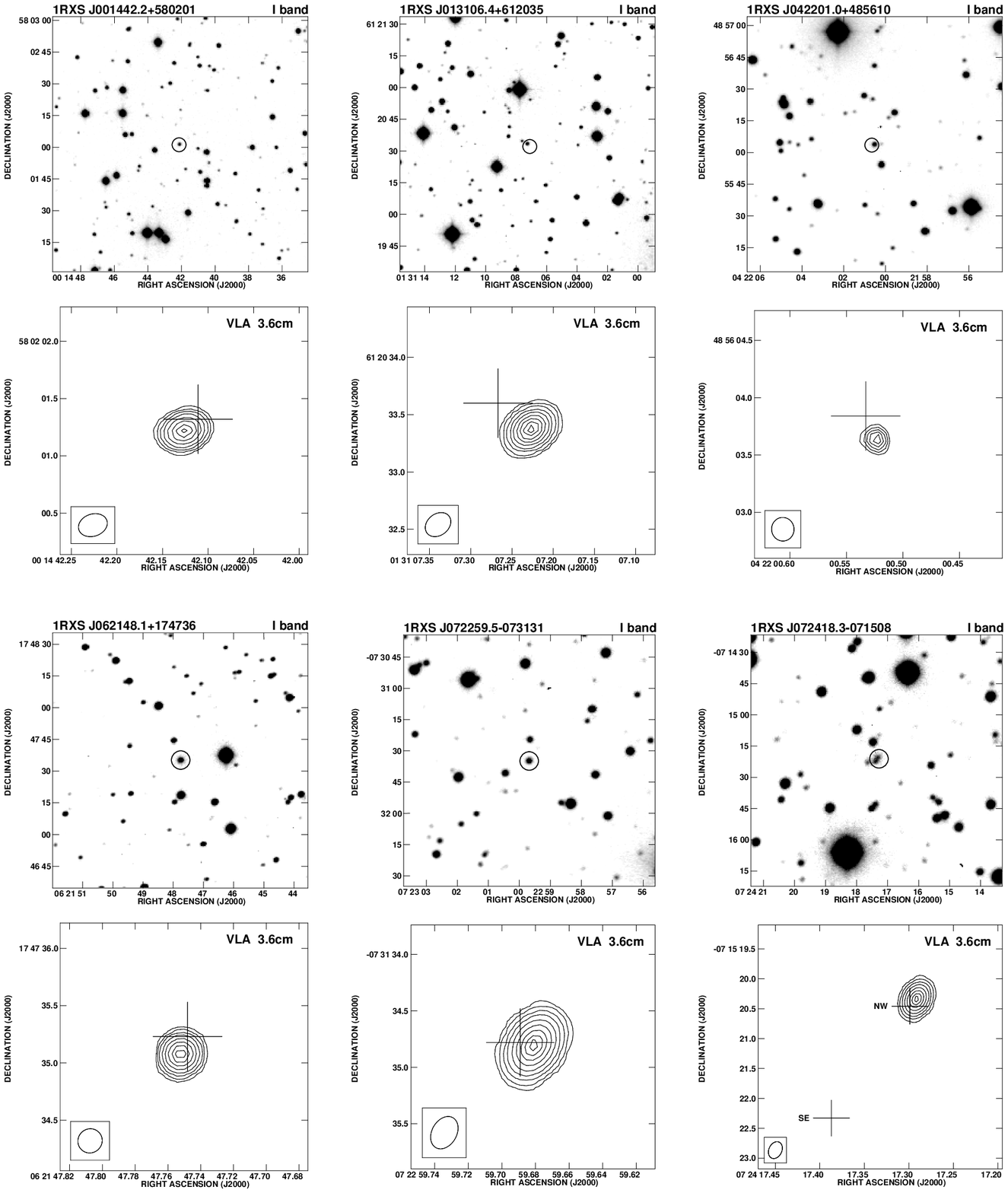}}
\caption{Optical images and radio maps, grouped in pairs, of the six sources listed in Table~\ref{radiot}. The $2\arcmin\times2\arcmin$ optical images were obtained through the $I$ Johnson filter with the CAHA 2.2~m telescope on 1999 December 11, and circles have been used to clearly mark the optical counterparts. The radio maps, computed using uniform weights, are the result of concatenating, for each source, all VLA 1999 July observations at 3.6\,cm wavelength, while the crosses indicate the optical positions with the 1$\sigma$ $0.3\arcsec$ astrometric errors. The radio maps are $2\arcsec\times2\arcsec$ wide for all sources except for the last one, where a $4\arcsec\times4\arcsec$ map has been plotted to allow the inclusion of the close SE optical object. The synthesized beam is plotted in the bottom left corner of each map. The radio contours start at a signal-to-noise ratio of 6 for all sources except for the last one, where a value of 15 has been used, in order to avoid reproducing cleaning effects at low signal-to-noise ratios in all cases.}
\label{optic_radio}
\end{figure*}

In order to obtain good quality maps, once accurate positions were known, we
carried out a similar process, but using phase self-calibration with the
previously obtained positions for each individual snapshot at 3.6~cm
wavelength. This was not possible for \object{1RXS~J042201.0+485610} because
the flux density was too low and prevented self-calibration. The resulting
uniform weighted maps, together with the optical images that will be presented
in the following section, are shown in Fig.~\ref{optic_radio}.

\begin{figure}[htpb]
\center
\includegraphics[height=19.3cm]{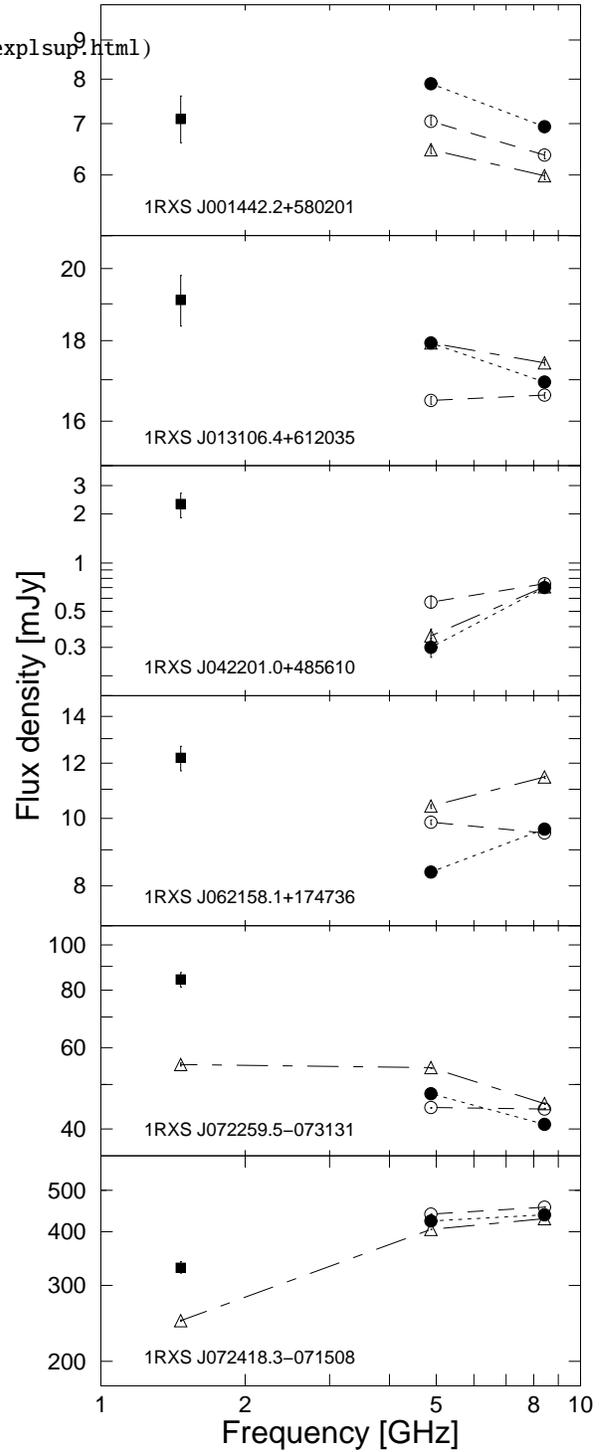}
\caption{Radio spectra of the six sources listed in Table~\ref{radiot} at different epochs. The VLA data from July 9, 22 and 30 are denoted by filled circles, open circles and open triangles, respectively. The non-simultaneous NVSS data, at 20~cm wavelength, are plotted as filled squares. Both axes are in logarithmic scale. Error bars not visible are smaller than the symbol's size.}
\label{radio_spectra}
\end{figure}

\begin{figure}[htpb]
\resizebox{\hsize}{!}{\includegraphics{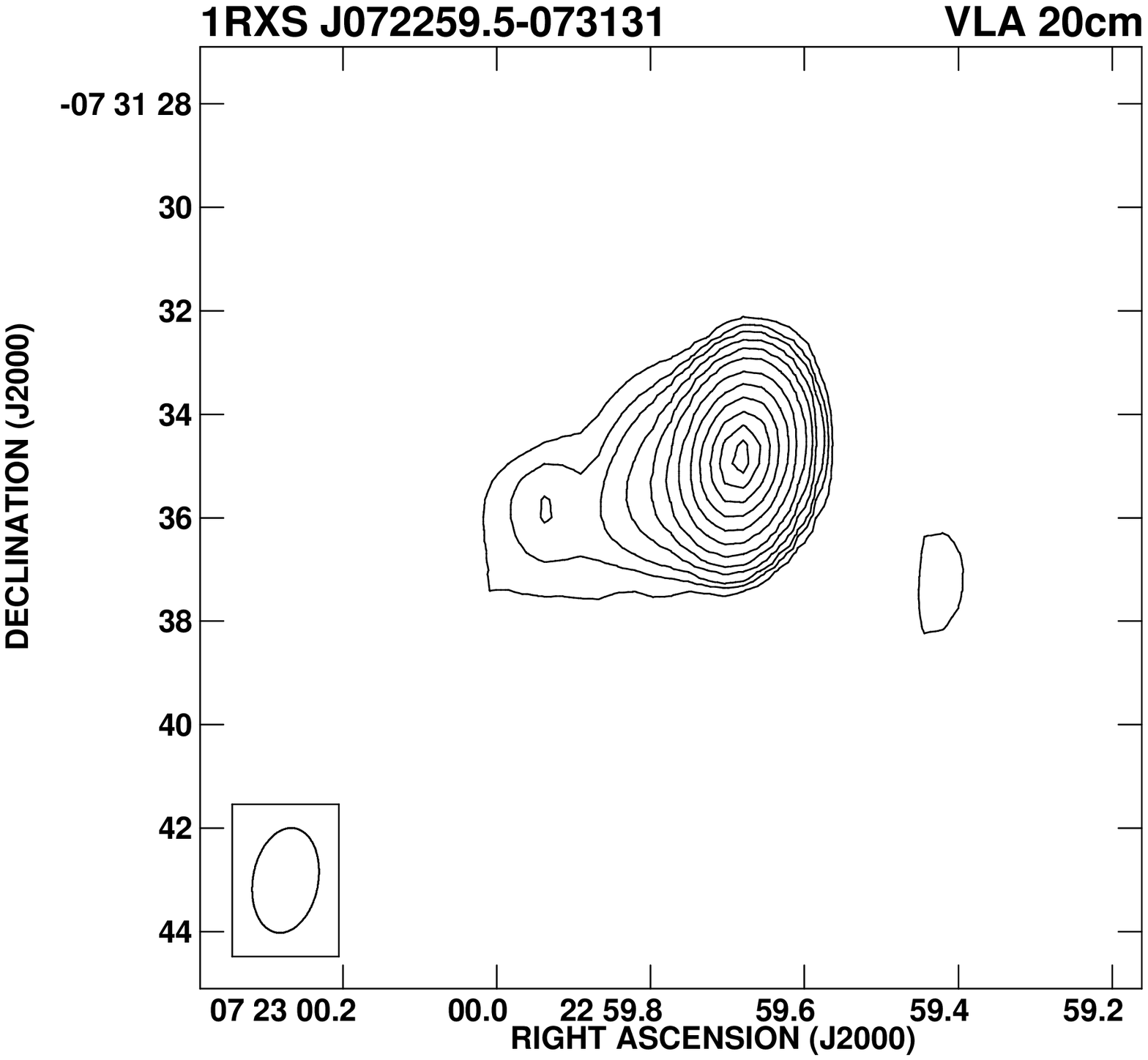}}
\caption{Natural weighted map of the self-calibrated 20~cm data of \object{1RXS~J072259.5$-$073131} obtained on the July 30 snapshot observation. Contours correspond to $-$3, 3, 4, 5, 7, 10, 15, 25, 40, 60, 85, 110 and 130 times 0.4~mJy~beam$^{-1}$, the rms noise.}
\label{jet}
\end{figure}

The flux densities of all sources were measured as follows. First of all we
self-calibrated all snapshots at all wavelengths (again with the exception of
\object{1RXS~J042201.0+485610}). Then we performed natural weighted maps with
improved dynamic range, from which flux densities were measured using again
elliptical Gaussian fits.

The obtained results for all observed sources are summarized in
Table~\ref{radiot}. In Col.~1 we list the RBSC object names, while in Cols.~2
and 3 we show the radio positions derived from our phase-referenced uniform
weighted maps of all concatenated 3.6~cm wavelength observations. In Col.~4 we
list the observing dates and, in Cols.~5 and 6, the corresponding flux
densities at both observing wavelengths, measured on the self-calibrated
natural weighted maps. The flux density errors quoted are the rms noise of the
maps in mJy beam$^{-1}$. In Col.~7 we show the computed spectral index
$\alpha$ (defined in such a way that $S_{\nu}\propto\nu^{+\alpha}$, where
$S_{\nu}$ is the flux density at a given frequency $\nu$) of each source for
all epochs. The radio spectra corresponding to the values listed in
Table~\ref{radiot} are plotted in Fig.~\ref{radio_spectra} for all sources,
including the 20~cm wavelength flux densities measured on July 30 for the last
two sources. We have also plotted, for all sources, the non-simultaneous 20~cm
measurements from the NVSS survey obtained in VLA D configuration.

Inspection of all obtained maps (with phase-reference or self-calibration
techniques, with uniform or natural weights, of individual snapshots or after
concatenating all of them, and at all frequencies), reveals that all sources
are point-like except \object{1RXS~J072259.5$-$073131}, which presents
extended structure towards the east, marginally present in the 3.6 and 6~cm
maps, and visible as a one-sided jet in the 30 July snapshot observation at
20~cm wavelength (see Fig.~\ref{jet}).

\section{Optical observations} \label{optical}

Having in mind the goal of identifying the optical counterparts, CCD
observations were made at Calar Alto (Almer\'{\i}a, Spain) with the 1.5~m
telescope of the Observatorio Astron\'omico Nacional (OAN), between 24 and 30
November 1998. We used the Ritchey-Chr\'etien focus and a Tektronics TK1024AB
chip that provides a scale factor of $0.4\arcsec$ per pixel and a
$6.9\arcmin\times6.9\arcmin$ field of view. Deep CCD images were obtained for
all our candidates, with the exception of \object{1RXS~J181119.4$-$275939} and
\object{1RXS~J185002.8$-$075833} that were not visible at the epoch of our
observations. The images were taken with the $V$, $R$ and $I$ Johnson filters,
and the exposure times were typically of 600--1200 seconds. Although the
images were useful for photometric measurements, their quality did not allow
us to obtain accurate positions.

Complementary CCD observations were conducted on 1999 December 11 also at
Calar Alto, but using the 2.2~m telescope of the Centro Astron\'omico
Hispano-Alem\'an (CAHA). Images were obtained using the Ritchey-Chr\'etien
focus and CAFOS, leading a scale factor of $0.53\arcsec$ per pixel and a
$8.8\arcmin\times8.8\arcmin$ field of view, through the $I$ Johnson filter and
with exposure times between 180 and 600 seconds. The good quality of these
images allowed us to perform accurate astrometric measurements, although no
precise photometric information could be obtained since the images could not
be photometrically calibrated.

The observations were reduced using standard procedures (bias and dark
subtraction and flat-field correction) within the IRAF software package.

\subsection{Results}

Promising optical counterparts for all observed sources, i.e., the first six
ones in Table~\ref{crosst}, were found in both observing runs. Thanks to the
CAHA 2.2~m observations we were able to obtain accurate positions. For this
purpose, we performed a detailed astrometric reduction of each image using
field stars present in the USNO-A2.0 catalog (Monet et~al. \cite{monet99}).
First of all we selected point-like objects in the image, not elliptical or
binary objects due to crowded fields, and determined their positions ($X$,
$Y$). We also rejected very faint objects to allow a significant
signal-to-noise ratio. On the other hand, very bright objects were also
rejected, because the positions given in the USNO-A2.0 catalog are from plates
obtained around 1950, and nearby (bright) stars could have experienced a shift
in position since then due to relatively high proper motions. Then we fitted
an astrometric solution to our image using the USNO-A2.0 ($\alpha$, $\delta$)
and image ($X$, $Y$) positions for the common stars in both the catalog and
the image, and rejected spurious points above 3$\sigma$. We proceeded
iteratively until convergence was achieved. This process allowed us to obtain
plate solutions with an rms of $\sim0.2\arcsec$ in each coordinate. This
relatively high error probably arises from the non-zero proper motions of the
field stars finally used in the fit, between 36 and 155 depending on each
image. We have to quadratically add to this error the 1$\sigma$ empirical
uncertainty estimate of the USNO-A2.0 catalog relative to the ICRF, which is
also $\sim0.2\arcsec$ in each coordinate (see Table~1 of Deutsch
\cite{deutsch99}). Therefore, the final 1$\sigma$ error of our obtained
coordinates is estimated to be 0.3$\arcsec$.

We show the central $2\arcmin\times2\arcmin$ of the CAHA 2.2~m images in
Fig.~\ref{optic_radio}, with the optical counterparts marked with a circle
(which is not any X-ray or radio error box). All sources appeared basically as
point-like objects, except the quasar \object{1RXS~J072418.3$-$071508}, which
had a complex structure with two separate optical components, hereafter NW
(northwest) and SE (southeast). The NW position falls well within the NVSS
error box and close to our obtained radio position. On the contrary, the SE
position is clearly out the NVSS error box, and cannot be considered as a
reliable counterpart to the radio source.

Photometry was obtained thanks to the OAN 1.5~m observations. The absolute
photometry is believed to be accurate only to $\pm$0.1 magnitude in the best
cases, because we could only determine the approximate photometric zero point
by observing a few standard stars from Landolt (\cite{landolt92}).

The obtained results are presented in Table~\ref{opticalt}, where Col.~1 gives
the RBSC object name, Cols.~2 and 3 give the J2000.0 ICRS coordinates of the
optical position with the 1$\sigma$ error of 0.3$\arcsec$ below each
coordinate, Col.~4 gives the observing dates and Cols.~5 to 7 give the Johnson
$V$, $R$ and $I$ magnitudes. Magnitudes for \object{1RXS~J072418.3$-$071508}
correspond to the sum of both the NW and SE components.

\section{Discussion} \label{discussion}

The observations reported here provide accurate positions in the optical
(0.3$\arcsec$) and specially in the radio (0.01$\arcsec$) domains. The
astrometric agreement for each source can be seen in the radio maps of
Fig.~\ref{optic_radio}, and is numerically expressed in Table~\ref{offsetst},
where we list the offsets in right ascension, in declination and the total
offset between the optical and radio positions. As can be seen, the offsets in
each coordinate are always smaller than the 1$\sigma$ optical errors in
position.

Although the agreements in position are certainly very promising, we have
estimated the probability of a random coincidence. For this purpose, we have
obtained the limiting $I$ magnitude of our CAHA 2.2~m images and counted the
number of objects until this threshold. Then we have estimated the
object-density of our images by assigning to each object the area of the
1$\sigma$ error box in position, i.e., $(2\times1\sigma)^2=(0.6\arcsec)^2$.
Finally, we have divided the area occupied by the stars by the area of the
image, and obtained a naive estimate of the probability to find an optical
object, up to a given limiting $I$ magnitude, close to the radio position by
less than 1$\sigma$, or 0.3$\arcsec$, in each coordinate. The limiting $I$
magnitudes and obtained probabilities of random coincidence are listed in the
last two columns of Table~\ref{offsetst}. Since these probabilities are always
smaller than 1\%, and even much smaller if we consider the particular offsets
instead of the 1$\sigma$ optical errors, we conclude that probably all optical
objects are the counterparts of the radio sources, and none of them is a field
star not related to the radio source.

On the other hand, an analysis of observed radio spectra of X-ray binaries can
be found in Fender (\cite{fender01}). It seems clear that negative spectral
indices ($-1\le\alpha\le-0.2$) are detected when observing synchrotron
emission from expanding plasmons, which are the result of discrete ejections
after an outburst. On the contrary, flat or inverted radio spectra
($0.0\le\alpha\le0.6$) are typical of the low/hard X-ray state of black hole
candidates, and are believed to arise in synchrotron emission from a partially
self-absorbed jet. Hence, a variety of spectral indices can be found in the
objects we are looking for. However, inverted spectra with $\alpha>1$ at high
radio frequencies (above 5~GHz) would probably rule out a REXB nature.

REXBs often display a variable flux, although extragalactic sources may vary
as well. Our observations, performed at one/two week interval, allow us to
estimate the degree of variability given by $(S_{\rm max}-S_{\rm min})/(S_{\rm
max}+S_{\rm min})$, being $S_{\rm max}$ and $S_{\rm min}$ the maximum and the
minimum flux density, respectively. As it is clear from Table~\ref{radiot} and
Fig.~\ref{radio_spectra}, all sources in the sample show some degree of flux
density variability at centimeter wavelengths during the three runs of our
monitoring. This variability is always within 10\%, except for
\object{1RXS~J042201.0+485610} at 6~cm, where a variability up to $\sim30$\%
is detected, although it should be considered with caution due to the low
emission level.

The optical emission seen in microquasars arises from the non-degenerated star
of the system, i.e., the companion of the compact object. Hence, in
microquasars containing a high mass companion spectral types O or B are found,
while in those containing a low mass companion the spectral type is later than
A (White et~al. \cite{white95}). Therefore, as optical counterpart of a given
candidate, we expect to find a non-degenerated star with any spectral type,
and with a luminosity class ranging from the main sequence to supergiant. In
fact, once photometric magnitudes of an object through different filters and
its distance are known, one can easily deduce its spectral type assuming it is
a star. However, we have no information about the distance to our sources, and
no correction for extinction can be applied to our data. This is particularly
important in our case, since our search has been carried out at low galactic
latitudes, where a high degree of extinction is expected, preventing any
spectral type classification from the photometric data alone.

\begin{table*}
\begin{center}
\caption[]{Optical astrometric and photometric results for all sources listed in Table~\ref{radiot}. Accurate positions were obtained from the $I$ Johnson filter CAHA 2.2~m telescope observations carried out on 1999 December 11. Magnitudes in $V$, $R$ and $I$ Johnson filters were obtained in different dates of November 1998 using the OAN 1.5~m telescope. Note that two positions are given for the source \object{1RXS~J072418.3$-$071508}.}
\label{opticalt}
\begin{tabular}{lrrcccc}
\hline \hline \noalign{\smallskip}
1RXS name           & $\alpha$~(2000) & $\delta$~(2000) & Day & \multicolumn{3}{c}{Magnitude}\\
                    & [h, m, s]    & [$\degr$, $\arcmin$, $\arcsec$] & Nov. 1998 & $V$ & $R$ & $I$\\
\noalign{\smallskip} \hline \noalign{\smallskip}
J001442.2+580201    &       00 14 42.111 &   +58 02 01.32 & 25 & -            & -            & $>19.0$\\
                    &       $\pm\,0.038$ &    $\pm\,0.30$ & 27 & -            & -            & $20.1\pm0.4$\\
                    &                    &                & 27 & -            & -            & $19.7\pm0.2$\\
\noalign{\smallskip}
J013106.4+612035    &       01 31 07.267 &   +61 20 33.60 & 27 & $19.5\pm0.1$ & $18.8\pm0.1$ & $17.9\pm0.1$\\
                    &       $\pm\,0.042$ &    $\pm\,0.30$ &    &              &              &             \\
\noalign{\smallskip}
J042201.0+485610    &       04 22 00.533 &   +48 56 03.84 & 28 & $20.4\pm0.2$ & $18.8\pm0.1$ & $17.4\pm0.1$\\
                    &       $\pm\,0.030$ &    $\pm\,0.30$ & 29 & -            & -            & $17.5\pm0.1$\\
\noalign{\smallskip}
J062148.1+174736    &       06 21 47.748 &   +17 47 35.23 & 28 & $19.8\pm0.1$ & $18.6\pm0.1$ & $17.5\pm0.1$\\
                    &       $\pm\,0.021$ &    $\pm\,0.30$ & 29 & -            & -            & $17.6\pm0.1$\\
\noalign{\smallskip}
J072259.5$-$073131  &       07 22 59.689 & $-$07 31 34.78 & 28 & $18.5\pm0.1$ & -            & -           \\
                    &       $\pm\,0.020$ &    $\pm\,0.30$ & 29 & $18.3\pm0.1$ & $17.7\pm0.1$ & $16.8\pm0.1$\\
\noalign{\smallskip}
J072418.3$-$071508  & (NW)~~07 24 17.299 & $-$07 15 20.46 & 29 & $18.8\pm0.1$ & $18.0\pm0.1$ & $17.2\pm0.1$\\
                    &       $\pm\,0.020$ &    $\pm\,0.30$ &    &              &              &             \\
                    & (SE)~~07 24 17.387 & $-$07 15 22.33 &    &              &              &             \\
                    &       $\pm\,0.020$ &    $\pm\,0.30$ &    &              &              &             \\
\noalign{\smallskip} \hline \noalign{\smallskip}
\end{tabular}
\end{center}
\end{table*}

\begin{table*}
\begin{center}
\caption[]{Offsets from the radio to the optical positions in right ascension, in declination and the total offset. The limiting $I$ magnitude of our CAHA 2.2~m images and the probability of a random coincidence are listed in the last two columns.}
\label{offsetst}
\begin{tabular}{lccccc}
\hline \hline \noalign{\smallskip}
1RXS name           & $\alpha_{\rm o}-\alpha_{\rm r}$ [$\arcsec$] & $\delta_{\rm o}-\delta_{\rm r}$ [$\arcsec$] & o-r [$\arcsec$] & Limiting $I$ mag. & Probability [\%]\\
\noalign{\smallskip} \hline \noalign{\smallskip}
J001442.2+580201    &   $-$0.12  &  \,~~0.10  &  0.16  &  22  &  0.3 \\
J013106.4+612035    &  \,~~0.29  &  \,~~0.22  &  0.37  &  21  &  0.3 \\
J042201.0+485610    &  \,~~0.08  &  \,~~0.21  &  0.22  &  20  &  0.2 \\
J062148.1+174736    &   $-$0.06  &  \,~~0.15  &  0.16  &  20  &  0.2 \\
J072259.5$-$073131  &  \,~~0.12  &  \,~~0.02  &  0.12  &  19  &  0.2 \\
J072418.3$-$071508  &  \,~~0.12  &   $-$0.12  &  0.17  &  18  &  0.2 \\
\noalign{\smallskip} \hline
\end{tabular}
\end{center}
\end{table*}

\subsection{Discussion on individual sources} \label{individual}

\paragraph{\object{1RXS~J001442.2+580201}.} The radio counterpart, see
Table~\ref{radiot}, shows a moderate degree of variability but always
displaying a negative spectral index of $\sim-0.2$, suggesting an optically
thin or partially self-absorbed synchrotron emitter. As can be seen in
Fig.~\ref{radio_spectra}, the non-simultaneous NVSS flux density measurement
at 20~cm wavelength is compatible with this behavior or even with a flattening
of the spectrum at longer wavelengths, although this could be due to intrinsic
variability. Non-thermal synchrotron radiation remains therefore as the most
plausible interpretation for the radio emission of this source. The optical
counterpart to the radio source (Fig.~\ref{optic_radio}) appears as a
point-like source with $I\sim20$, which makes it the weakest optical object of
our sample. No $R$ or $V$ magnitudes could be obtained, although a typical
behavior for low latitude highly absorbed sources is an increase of $\sim1$
magnitude when changing from $I$ to $R$ and from $R$ to $V$, as can be seen in
the other sources of Table~\ref{opticalt}. All the available information
points towards a good REXB candidate.

\paragraph{\object{1RXS~J013106.4+612035}.} In the radio domain it shows a
slightly negative or close to zero spectral index, reminiscent of optically
thin or partially self-absorbed jets in REXBs, in which the non-simultaneous
NVSS flux measurement fits well, and it displays a very low degree of
variability of 2\% at 3.6~cm and 4\% at 6~cm. This object has also been
observed within the Green Bank 5~GHz radio survey (Gregory \& Condon
\cite{gregory91}; Gregory et~al. \cite{gregory96}) with a measured flux
density of 24~mJy. This value is a little bit higher than the ones reported
here, although this discrepancy could be due to intrinsic variability and/or
to the different angular scales sampled by the interferometric VLA
observations. In fact, VLA A configuration observations carried out by
Laurent-Muehleisen et~al. (\cite{laurent97}) on October 19, 1992, indicate a
flux density of 16~mJy at 5~GHz, in good agreement with our data. In the
optical domain, the counterpart to the radio source appears as a point-like
object with a nearby field object at $\sim3\arcsec$ to the northeast, visible
in Fig.~\ref{optic_radio}, which is $\sim1$ magnitude fainter in the $I$ band.
The optical magnitudes in Table~\ref{opticalt} correspond to the sum of both
objects (unresolved in the OAN 1.5~m images), and are typical of low latitude
highly absorbed sources. Overall it seems a good REXB candidate.

\paragraph{\object{1RXS~J042201.0+485610}.} In the RBSC this source appears
catalogued with an X-ray extension of $17\arcsec$, which could suggest an
extragalactic origin of the source. However, X-ray halos produced by the
scattering of the original X-ray photons by interstellar dust grains are seen
in some galactic microquasars, like for example in a Chandra observation of
\object{Cyg~X-3} (Predehl et~al. \cite{predehl00}). In fact, \object{Cyg~X-3}
has an X-ray extension of $63\arcsec$ within the RBSC, while \object{SS~433}
has one of $25\arcsec$. Hence, the presence of an X-ray extension does not
rule out a REXB nature for this source. The radio counterpart shows the lowest
radio emission level of all sources in our sample, and displays an inverted
spectrum up to $\alpha\sim+1.6$ (see Table~\ref{radiot}), difficult to account
for self-absorbed synchrotron radiation at such high frequencies.
Nevertheless, as it happens in some AGN, a highly inverted spectrum would be
expected if the jet is stopped in a dense medium on a compact scale. On the
other hand, the non-simultaneous NVSS flux density measurement seems to be
incompatible with self-absorbed synchrotron radiation. Although this could be
due to intrinsic variability of the source, this would require variations of
an order of magnitude in flux. This discrepancy can be explained if we assume
a thermal origin for the radio emission and if we attribute the drop in flux
density in our VLA A configuration observations due to over resolution when
compared to the NVSS VLA D configuration observations. The optical counterpart
to the radio source has a nearby field object at $\sim3\arcsec$ to the west,
marginally visible in Fig.~\ref{optic_radio}, which is $\sim2$ magnitudes
fainter in the $I$ band. The optical magnitudes in Table~\ref{opticalt}
correspond to the sum of both objects (unresolved in the OAN 1.5~m images),
although the nearby one is not expected to contaminate appreciably, and are
again typical of low latitude highly absorbed sources. Although at first sight
our target appears as a point-like source in Fig.~\ref{optic_radio}, if we
perform Gaussian fits to field objects with similar magnitudes, we find our
target to have a Full Width at Half Maximum (FWHM) $\sim15$\% greater,
suggesting that it is in fact an extended optical object. In summary, the high
spectral index and the extended optical emission indicate that this source is
not a very promising REXB candidate.

\paragraph{\object{1RXS~J062148.1+174736}.} In the radio domain, this source
displays an inverted or flat spectrum, which could account for a partially
self-absorbed jet. On the other hand, the moderate variability observed,
around 10\%, could explain the higher non-simultaneous NVSS flux density
measurement.  In the optical domain, this source appears as a point-like
object with magnitudes typical of low latitude highly absorbed sources.
However, a careful inspection of the optical image, reveals that the FWHM of a
Gaussian fit to our target source is $\sim30$\% greater than those of other
sources with similar magnitudes. This fact points to an extragalactic nature
for this source. It is worth to mention that this source is the only one in
our sample present in the second incremental release of the Two Micron All Sky
Survey (2MASS) (Cutri et~al. \cite{cutri00}). Its position, obtained from
images taken in October 1997, is there listed as $\alpha=6^{\rm h}~21^{\rm
m}~47.749^{\rm s}\pm0.009^{\rm s}$,
$\delta=+17\degr~47\arcmin~35.07\arcsec\pm0.13\arcsec$, in very good agreement
with our obtained coordinates. Finally, we would like to point out that Motch
et~al. (\cite{motch98}) list 3 stars within the X-ray error box as possible
counterparts to the X-ray source. Our proposed optical counterpart on the
basis of the radio position, would rule out any of these 3 stars as the
optical counterpart of this X-ray source. Overall, it looks like an
extragalactic object because of its optical extension.

\paragraph{\object{1RXS~J072259.5$-$073131}.} The radio counterpart presents
an extended structure towards the east, marginally present in the 3.6 and 6~cm
maps, and visible as a one-sided jet in the 30 July 4 minute snapshot
observation at 20~cm wavelength (see Fig.~\ref{jet}). Although the one-sided
morphology is suggestive of an extragalactic nature for this source, specially
if we take into account that we are mapping the source at arcsecond scales, we
cannot rule out a possible galactic nature on the basis of the detected
morphology. The source also shows a moderate degree of variability, displaying
most of time a negative spectral index but also close to 0 in the July 22
observation. As can be seen in Fig.~\ref{radio_spectra}, the non-simultaneous
NVSS flux density measurement at 20~cm wavelength could be compatible with a
non-thermal optically thin spectrum. However, simultaneous observations during
July 30 show a flattening of the spectrum at higher wavelengths, which could
be due to self-absorption in this energy range. The difference between the
NVSS and our flux density measurement at 20~cm wavelength could be due to
intrinsic variability and/or to the different VLA configurations used. Anyway,
non-thermal synchrotron radiation remains as the most plausible interpretation
for the radio emission of this source, specially at higher frequencies. The
optical counterpart appears as a point-like object with the magnitudes of an
absorbed object at low galactic latitudes. These properties are in good
agreement with the expected ones for a microquasar, although the one-sided
radio jet at arcsecond scales is most common in extragalactic objects.

\paragraph{\object{1RXS~J072418.3$-$071508}.} This source has been recently
(March 2002) classified as a quasar in the SIMBAD database, and it is not any
more a REXB candidate. It is listed as \object{PMN~J0724$-$0715} in the NED
database, and is the source \object{WGA~J0724.3$-$0715} in Perlman et~al.
(\cite{perlman98}), who reported a faint and quite broad H$\alpha$ emission
line (rest-frame $W_{\lambda}=30.3$~\AA, FWHM=4000~km~s$^{-1}$), and
classified it as a flat spectrum radio quasar with $z=0.270$. Nevertheless, we
have reported here our observational results for this source, since it was a
candidate when we performed the observations. In the radio it displays a low
degree of variability of 3\% at 3.6~cm and 4\% at 6~cm, with a flat or even
inverted spectrum. In the optical it appears as a complex double object, with
a NW and a SE components, being the first one the counterpart to the radio
source. Magnitudes in $V$, $R$ and $I$ typical of absorbed objects at low
galactic latitudes were found. On the other hand, it appears catalogued with
an X-ray extension of $40\arcsec$ in the RBSC.

\section{Conclusions} \label{conclusions}

We have presented a cross-identification method to search for REXBs, which are
potential microquasar sources. The obtained results give confidence to the
proposed method, since the output list of objects included all but one known
persistent HMXB microquasars within $|b|<5\degr$. The unidentified sources in
the list were divided in two groups, depending on the offset between the RBSC
and NVSS positions. The Group~1 source list contained 8 objects, of which
\object{1RXS~J181119.4$-$275939} and \object{1RXS~J185002.8$-$075833} were not
studied because they were not visible during our observations. We studied the
remaining 6 radio sources of Group~1, and found optical counterparts to all of
them. We obtained accurate positions at both radio and optical wavelengths,
perfectly compatible between them. We also obtained radio spectra and optical
magnitudes of the sources.

After a detailed analysis of the obtained data, we conclude that
\object{1RXS~J001442.2+580201} and \object{1RXS~J013106.4+612035} are good
REXB candidates, while \object{1RXS~J042201.0+485610} is not so promising due
to its highly inverted spectrum at high frequencies and its optical extended
emission. A careful study of the optical counterpart of
\object{1RXS~J062148.1+174736} reveals it is extended, which points towards an
extragalactic nature for this object. The situation is not clear in the case
of \object{1RXS~J072259.5$-$073131}, because the optical data agrees with a
microquasar nature, while the detected one-sided radio jet reaching arcsecond
scales is most common in extragalactic sources. The last studied source,
\object{1RXS~J072418.3$-$071508}, is an already identified quasar. Hence, two
sources in our sample are very promising to be new microquasars in the Galaxy.

Optical spectroscopic observations, intended to unveil the nature of the
optical counterparts, were attempted in autumn 2000 and 2001, and were not
carried out due to bad weather conditions. A new attempt is in progress for
autumn 2002. On the other hand, VLBI observations of these sources will be
reported soon (Rib\'o et~al. \cite{ribo02a}).

The two sources belonging to Group~1 not yet observed, namely
\object{1RXS~J181119.4$-$275939} and \object{1RXS~J185002.8$-$075833} will be
eventually studied in the future, together with Group~2 sources. Finally,
since some microquasars could have been ejected from the galactic plane
(Rib\'o et~al. \cite{ribo02b}), we are planning to extend this project up to
higher galactic latitudes ($5\degr<|b|<10\degr$).

\begin{acknowledgements}

We acknowledge useful comments and discussions with O. Fors, X. Otazu and I. Ribas.
We also acknowledge J.~S. Bloom for reading through a draft version of this work and for his useful suggestions.
We acknowledge detailed and useful comments from H. Falcke, the referee of this paper.
We acknowledge partial support by DGI of the Ministerio de Ciencia y Tecnolog\'{\i}a (Spain) under grant AYA2001-3092.
We also acknowledge partial support by the European Regional Development Fund (ERDF/FEDER).
M.~R. is supported by a fellowship from CIRIT (Generalitat de Catalunya, ref. 1999~FI~00199).
J.~M. is partially supported by Junta de Andaluc\'{\i}a (Spain) and has also been aided in this work by an Henri Chr\'etien International Research Grant administered by the American Astronomical Society.
The authors were Visiting Astronomers at German-Spanish Astronomical Centre, Calar Alto, operated by the Max-Planck-Institute for Astronomy, Heidelberg, jointly with the Spanish National Commission for Astronomy.
This research has made use of the NASA's Astrophysics Data System Abstract
Service, of the SIMBAD database, operated at CDS, Strasbourg, France, and of
the NASA/IPAC Extragalactic Database (NED) which is operated by the Jet
Propulsion Laboratory, California Institute of Technology, under contract with
the National Aeronautics and Space Administration. 
The Digitized Sky Survey was produced at the Space Telescope Science Institute under U.S. Government grant NAG~W-2166.

\end{acknowledgements}


\begin{thebibliography}{}

\bibitem[2001]{castro01}
Castro-Tirado, A.~J., Greiner, J., \& Paredes, J.~M.
2001, Proc. of the Third Microquasar Workshop on `Galactic Relativistic Jet Sources', ed. A.~J. Castro-Tirado, J. Greiner, \& J.~M. Paredes (Kluwer Academic Publishers), Ap\&SS, 276

\bibitem[1998]{condon98}
Condon, J.~J., Cotton, W.~D., Greisen, E.~W., et~al.
1998, AJ, 115, 1693

\bibitem[2000]{cutri00}
Cutri, R.~M., Skrutskie, M.~F., Van Dyk, S., et~al. 
2000, {\it Explanatory Supplement to the 2MASS, Second Incremental Data Release}, Caltech (available at {\tt http://www.ipac.caltech.edu/2mass/releases/second/\\doc/explsup.html})

\bibitem[1999]{deutsch99}
Deutsch, E.~W.
1999, AJ, 118, 1882

\bibitem[2000]{fender00}
Fender, R.~P., \& Hendry, M.~A.
2000, MNRAS, 317, 1

\bibitem[2001]{fender01}
Fender, R.~P.
2001, MNRAS, 322, 31

\bibitem[2001]{fomalont01}
Fomalont, E.~B., Geldzahler, B.~J., \& Bradshaw, C.~F.
2001, ApJ, 558, 283

\bibitem[1991]{gregory91}
Gregory, P.~C., \& Condon, J.~J.
1991, ApJS, 75, 1011

\bibitem[1996]{gregory96}
Gregory, P.~C., Scott, W.~K., Douglas, K., \& Condon, J.~J.
1996, ApJS, 103, 427

\bibitem[2001]{hannikainen01}
Hannikainen, D., Campbell-Wilson, D., Hunstead, R., et~al.
2001, in Proc. of the Third Microquasar Workshop `Galactic Relativistic Jet Sources', ed. A.~J. Castro-Tirado, J. Greiner, \& J.~M. Paredes (Kluwer Academic Publishers), Ap\&SS, 276, 45

\bibitem[1999]{hartman99}
Hartman, R.~C., Bertsch, D.~L., Bloom, S.~D., et~al.
1999, ApJS, 123, 79

\bibitem[1997]{kniffen97}
Kniffen, D.~A., Alberts, W.~C.~K., Bertsch, D.~L., et~al.
1997, ApJ, 486, 126

\bibitem[1992]{landolt92}
Landolt, A.U.
1992, AJ, 104, 340

\bibitem[1997]{laurent97}
Laurent-Muehleisen, S.~A., Kollgaard, R.~I., Ryan, P.~J., et~al.
1997, A\&AS, 122, 235

\bibitem[2000]{liu00}
Liu, Q.~Z., van Paradijs, J., \& van den Heuvel, E.~P.~J.
2000, A\&AS, 147, 25

\bibitem[2001]{liu01}
Liu, Q.~Z., van Paradijs, J., \& van den Heuvel, E.~P.~J.
2001, A\&A, 368, 1021

\bibitem[2001]{marti01}
Mart\'{\i}, J., Paredes, J.~M., \& Peracaula, M.
2001, A\&A, 375, 476

\bibitem[2001]{massi01}
Massi, M., Rib\'o, M., Paredes, J.~M., Peracaula, M., \& Estalella, R.
2001, A\&A, 376, 217

\bibitem[1999]{mirabel99}
Mirabel, I.~F., \& Rodr\'{\i}guez, L.~F.
1999, ARA\&A, 37, 409

\bibitem[1999]{monet99}
Monet, D.~G., Bird, A., Canzian, B., et~al.
1999, USNO-A2.0 CD-ROM (U.S. Naval Observatory, Washington DC)

\bibitem[1998]{motch98}
Motch, C., Guillout, P., Haberl, F., et~al.
1998, A\&AS, 132, 341

\bibitem[2000]{paredes00}
Paredes, J.~M., Mart\'{\i}, J., Rib\'o, M., \& Massi, M.
2000, Science, 288, 2340

\bibitem[1998]{perlman98}
Perlman, E.~S., Padovani, P., Giommi, P., et~al.
1998, AJ, 115, 1253

\bibitem[2000]{predehl00}
Predehl, P., Burwitz, V., Paerels, F., \& Tr\"umper, J.
2000, A\&A, 357, L25

\bibitem[2002a]{ribo02a}
Rib\'o, M., Ros, E., Paredes, J.~M., Massi, M., \& Mart\'{\i}, J.
2002a, A\&A, in press

\bibitem[2002b]{ribo02b}
Rib\'o, M., Paredes, J.~M., Romero, G.~E., et~al.
2002b, A\&A, 384, 954

\bibitem[2001]{rodriguez01}
Rodr\'{\i}guez, L.~F., \& Mirabel, I.~F.
2001, in Proc. of The Nature of Unidentified Galactic High-energy Gamma-ray Sources, ed. A. Carrami\~nana, O. Reimer, \& D.~J. Thompson,
Astrophysics and Space Science Library, vol. 267, 245

\bibitem[2001]{stirling01}
Stirling, A.~M., Spencer, R.~E., de la Force, C.~J., et~al.
2001, MNRAS, 327, 1273

\bibitem[1996]{voges96}
Voges, W., Boller, Th., Dennerl, K., et~al.
1996, in Proc. of the Conference R\"ontgenstrahlung from the Universe, MPE Report, 263, 637

\bibitem[1999]{voges99}
Voges, W., Aschenbach, B., Boller, Th., et~al.
1999, A\&A, 349, 389

\bibitem[1995]{white95}
White, N.~E., Nagase, F., \& Parmar, A.~N.
1995, The properties of X-ray binaries, in X-ray Binaries, ed. W.~H.~G. Lewin, J. van Paradijs, \& E.~P.~J. van den Heuvel (Cambridge Univ. Press, Cambridge), 1

\end{thebibliography}
\end{document}